# Tapping opportunity of tiny shaped particles and role of precursor in developing shaped particles

**Mubarak Ali, [a], * and I -Nan Lin [b]**

[a] Department of Physics, COMSATS Institute of Information Technology, Islamabad 45550, Pakistan. *E-mail: mubarak74@comsats.edu.pk, mubarak74@mail.com

[b] Department of Physics, Tamkang University, Tamsui Dist., New Taipei City 25137, Taiwan (R.O.C.).

**Abstract** –Metallic colloids are frequently used in industry and provide understanding of science at microns to nanometers scales along with their applicability for various technologically important applications. Present investigations deal morphology-structure of gold, silver and their binary composition while processing the certain amount of their solutions in a newly designed process and tap opportunities of developing tiny shaped particles. At tuned ratio of pulse OFF to ON time and when gold solution was processed, several tiny shaped particles developed at solution surface. Such tiny particles deal force at the tip of each converted structure of smooth element under the steady-state immersing behavior pointing toward common centre to pack for developing different geometric anisotropic shaped particles. Under identical parameters along with pulse time, processing solutions of silver nitrate and binary composition of chloroauric acid-silver nitrate result into develop tiny particles having no specific shape where their packing deal mixed behavior of force resulting into develop distorted particles. Elongation and deformation of gold and silver atoms while in different structures is because of the plastically-driven behavior of electrons as per stretching of their clamped energy knots. In structure of three-dimensional where electrons of atoms do not deal transition require for elongation they retain the structure as it is known in hcp structure or two-dimensional structure. Different nature of precursors along with morphology-structure of particles is discussed in this paper opening abundant avenues of research.

**Introduction:**

Development of metallic colloids is simple, but it is crucial to understand the mechanisms of their formation. Development of different shaped particles demands alternative perspectives and explanations. It is possible to see features of a variety of materials at nanoscale as well as at atomic scale through high-resolution transmission microscopy. In certain nature precursor, different metallic elements may possess different affinity to counterparts, which may influence their dissociation rate. Again, due to different electronic configuration of atoms belonging to various elements their amalgamation rate can be different and in different manner. As a result, processing of various solutions in any scale of their composition may present pronounced effect in the formation of tiny particles, nanoparticles and particles related to targeted atoms. Then under the variation of processing approach where different parameters involved in developing materials of different features. Thus, to pinpoint reliable mechanisms of formation of tiny-sized particles (nanoscale components) following by their packing into different types of extended shapes, processing of relevant precursors under the same setup is essential.

Several approaches have been enlisted in the literature to process such metallic colloids, but citrate reduction method is the one of the most widely employed procedure [1]. Development of tiny metallic colloids and their likely coalescence into extended shapes have remained the subject of enormous studies and few of them are cited here [2-13]. Trapping of energetic electrons collectively oscillate the tiny lattice [2]. A tiny metal cluster is like simple chemical compound and may find several applications in catalysis, optics and electronics [3]. Nanocrystals have unique features and they have the tendency to extend in size providing options to fabricate advanced materials having better characteristics [4]. An ordered array of nanoparticles rather than agglomeration may give different properties of materials [5]. Coalescence of nanocrystals is a goal for developing particles [6]. Self-assembly means to design a specific structure [7]. Long-term goal of nanoparticle technology is to develop micro devices [8]. An initial effort is to assemble the nanoparticles [9]. Organization of nanometre size building blocks into specific structures is one of the current challenges [10]. Atoms and molecules will be treated as materials of tomorrow once they have fruitful assembling [11]. Complex shapes are possible by means of precise control on the assembling of nanoparticles [12]. Coalescence of nanocrystals into extended shapes provides endless choices [13].

Silver nanoparticles in narrow size distributions synthesized by employing direct laser irradiation method and the mechanism of their preparation were described [14]. Pentagon shapes nucleate at the cost of particles, which is kinetically unfavourable, [111] twin planes in silver and gold particles direct the shapes and it is due to re-entrant grooves, the existing mechanistic interpretations are insufficient to explain several observations and rate of reactant addition/reduction can be estimated to produce subsequent specific shaped particles in high yields [15]. Locating the specific mode of excitation of surface plasmon in metallic nanocrystals will bring intense consequences for the research fields [16]. Stress in the lattice may be due to hexagonal monolayer on 3-D surface or can be due to stacking faults [17]. More work is required to develop in-depth understanding of particles shapes and their structures in novel applications [18]. Suspended silver nanoparticles and their inter-particle spacing can be actively controlled in a microfluidic system [19]. While synthesizing silver nanoparticles in liquid pulse plasma, hydroxyl radical along with excited states of atoms, hydrogen and oxygen were detected in the emission spectra [20]. Observing an atom in high-resolution microscopy enables us to understand the functionalities of atoms [21].

Attempts have also been made to synthesize metallic particles in different plasma solution processes [22-29]. Gold nanoplates and nanorods are synthesized at the surface of solution while spherical-shaped particles are synthesized in the solution [22]. Under DC glow discharge, plasma irradiates the liquid which provides the main mechanisms of synthesis of crystals [24, 29].

Tiny particles of metallic colloids have shown tremendous potential for use as a catalyst in the new emerging applications of catalysis [30, 31] as phase-controlled syntheses give an improved catalytic activity of metal nanostructures than in bulk [32, 33]. The research efforts, in progress, should also consider dynamics in explaining the structure [34] and there are order metrics capable of accurately characterizing the order of packing [35].

Present study presents the development mechanisms of tiny-sized particles following by large-sized particles while processing molar concentrations of AgNO$_3$,

HAuCl$_4$.3H$_2$O+AgNO$_3$ and HAuCl$_4$.3H$_2$O precursors under the fixed ratio of pulse OFF to ON time. Morphology-structure of different metallic colloids is reported here, tapping the opportunity for tiny shaped particle; hence, their large size shaped particles are developed. This study pinpoints that under identical process parameters, the nature of the precursor takes the edge in terms of need-in atoms to develop a tiny shaped particle, and hence, a large shaped particle.

**Experimental details:**

Pulse DC power controller (SPIK2000A-20, MELEC GmbH Germany) was employed to generate bipolar pulse where ON/OFF time was set 10 μsec while processing the solutions of different precursors. In pulse-based electronphoton-solution interface setup, the bottom of the copper capillary was adjusted just above the surface of solution and layout of the pulse-based process is given elsewhere [36, 37].

Silver nitrate (AgNO$_3$) purity 6N was purchased from SIGMA-ALDRICH. The running voltage and current was recorded ~21 (V) and ~1.1 (A), respectively, which were controlled automatically once the light glow (known as plasma) was appeared on the splitting of flowing argon atoms and remained nearly constant throughout the process. Molar concentration of the silver precursor was 0.30 mM and was processed for 20 minutes at argon gas flow rate of ~ 100 sccm. In another experiment concentration of silver precursor was 0.60 mM and was processed for 40 minutes by maintaining argon gas flow rate 200 sccm in which the running voltage and current were recorded ~23 (V) and ~1.1 (A), respectively. In the case of binary composition (HAuCl$_4$: AgNO$_3$ = 75%:25%) where concentration of precursor was ~0.30 mM, the process time was 20 minutes, and argon gas flow rate was ~100 sccm, almost equal running voltage and current were recorded in processing silver precursor.

In the case of gold precursor, solid powder of HAuCl$_4$.3H$_2$O (Au 49.5 % min crystalline) was purchased from Alfa Aesar and different concentrations of solutions were prepared after mixing with DI water. An average running voltage and current were ~25 (V) and current ~1.2 (A), respectively. In processing gold solutions having concentration of precursor 0.20 mM, the process time was set to 5 minutes and 10 minutes where argon gas flow rate was ~100 sccm, whereas, in another experiment, precursor concentration was 0.60 mM which was processed for a duration of 5 minutes at argon gas having flow rate ~100 sccm. A slight fluctuation in the input power was recorded at the start of the process and the light glow sustained automatically in few seconds time in each experiment.

In each experiment, step-up transformer enhanced the voltage ~40 times. Temperature was measured by LASER-guided meter (CENTER, 350 Series) at the start, at the mid of process and at the end. Temperature of the processing solution was ~21°C at the start of the process, ~26°C at 5 minutes process time and ~34°C at 10 minutes process time. In the case of silver solution, the recorded value of the temperature was higher (~50°C for 20 minutes and ~65°C for 40 minutes). In each experiment, total amount of 100 ml solution was prepared.

To characterize and analyse gold, silver and their binary colloids, a drop of each prepared solution was poured on copper grid. Samples were dried in Photoplate degasser (JEOL EM-DSC30) for 24 hours to eliminate moisture. Bright field images of tiny-sized particles and large-sized particles were taken by transmission microscope known as TEM. High resolution images of tiny-sized particles and large-sized particles were collected by transmission microscope images known as HR-TEM (JEOL JEM2100F; 200 kV). Structural information was captured by selected area photon reflection (SAPR) pattern known as SAED pattern.

**Results and discussion:**

A large distorted silver particle is shown in Figure 1 (a) where different tiny particles packed under mixed behavior of force. High-resolution transmission microscope image taken from encircled region in Figure 1 (a) is shown in Figure 1 (b) where tiny particles possess different morphology-structure; tiny particles of less elongated atoms show different orientations (1), tiny particles pack by filling unfilled region completely (2), tiny particles pack by filling unfilled region moderately (3) and isolated tiny particle with no specific shape (4). Figure 1 (c) shows high-resolution view of transmission microscopy image, which is taken from the region covered under square box in Figure 1 (b) highlighting various structural aspects; a region reveals blurred surface (1), structure of different-level atoms elongated under the exertion of forces along their opposite poles (2), structure where all atoms elongated from the same ground level under the exertion of forces along their opposite poles (3), structure of elongated atoms introduced wrinkles under certain interaction of the medium (4), elongated atoms in a large region converted structure into structure of smooth elements (5), structure reveals a large region of blurred surface (6) and region where atoms don't deal compact configuration (7). A distorted silver particle is shown in Figure S1 (a) where no specific shape of tiny particle developed as is evident in its SAPR pattern, also, which is shown Figure S1 (A); intensity spots neither confined in precise dots nor retained order in the structure and distribution is uneven in the entire pattern. This reveals that there was no order in the structure even at moderate range and it is considered as the disordered structure in the form of tiny particle. In Figure S1 (b), distorted silver particles are shown in different size, which were developed under the packing of tiny particles other than triangular shape. Average size of particles is the same as in the case of Figure S1 (a) and pattern reveals chaotic structure as shown in Figure S1 (B). In Figures S2 (a) and (b), distorted particles of silver show different sizes and their average size is smaller as compared to the ones shown in Figure S1.

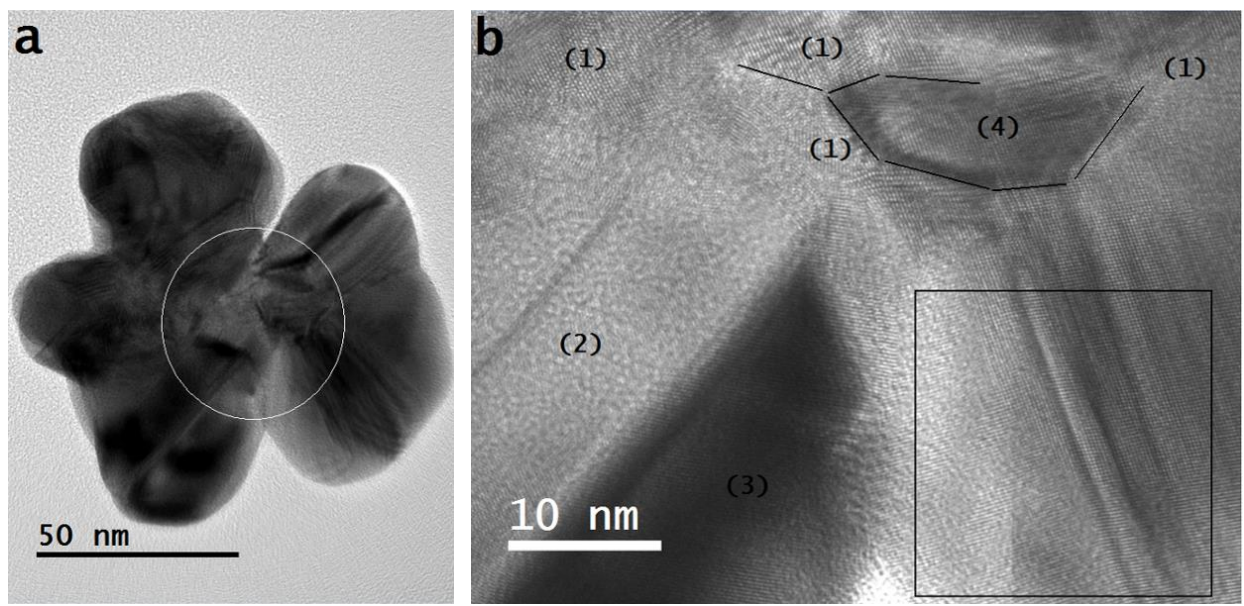

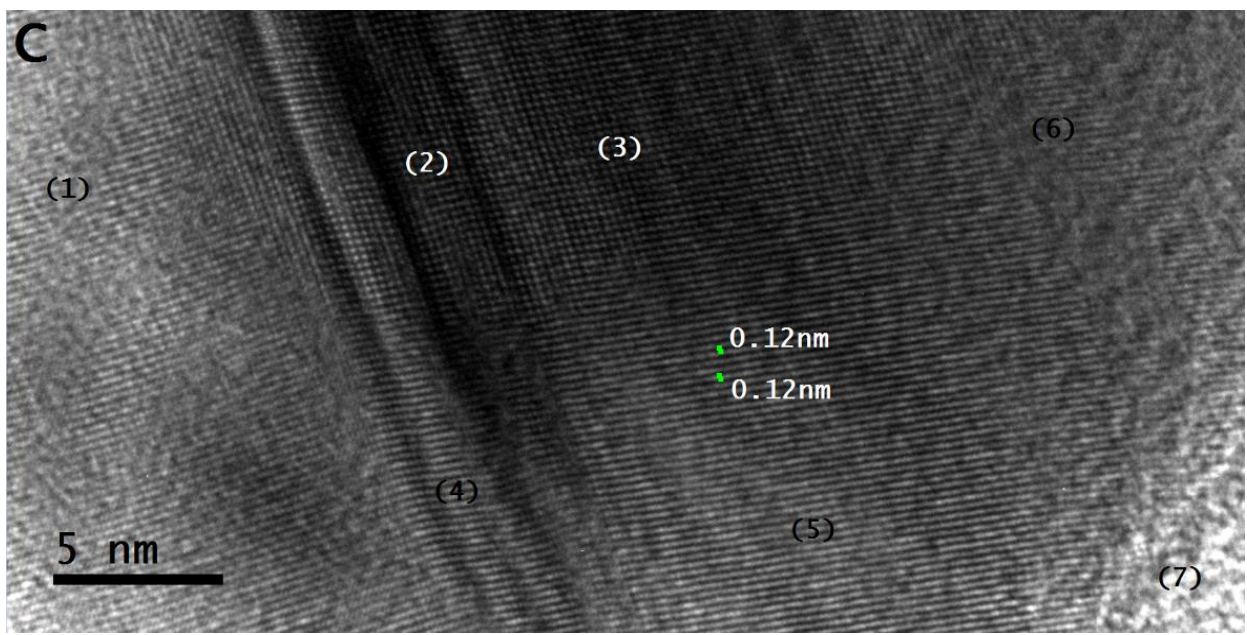

**Figure 1:** (a) Bright field transmission microscope image of silver distorted particle (left-side), (b) high-resolution transmission microscope image (right-side) of the encircled region in 'a' shows different structural aspects and (c) magnified high-resolution transmission microscope image of square region in 'b' further highlights different behaviors of atoms; precursor concentration: 0.30 mM, process duration: 20 minutes and argon gas flow rate: 100 sccm.

In another experiment, silver precursor was processed at higher molar concentration along with process time and argon gas flow rate. In Figure S3 (a-d), tiny particles possess no specific shape where distorted particles developed, which show the same morphological features as for those shown in Figure S2. In Figure 2 (a), high-resolution view of encircled region of Figure S3 (a) is shown, which indicates the elongation of atoms of one-dimensional arrays as in the case of Figure 1 (c) where surface defects are quite obvious at the boundaries of packing. High-resolution image taken from the encircled particle in Figure S3 (b) is shown in Figure 2 (b) where the composed structure reveals uniform stress in the region between the parallel drawn lines, region covered under the square box shows blurred surface while region covered under the rectangle shape doesn't show specific elongation of atoms. High-resolution transmission microscope image of the particle encircled in Figure S3 (c) is shown in Figure 2 (c) where entire surface is appeared to be blurred. In Figure 2 (d), elongated atoms of tiny particle reveal different structure; large region reveals blurred surface and region covered under rectangular box show elongation of atoms under force having tilted axis of exertion. Silver nanoparticles and particles shown in Figures S2 and S3 have the same morphological features as in the case of work given elsewhere [20].

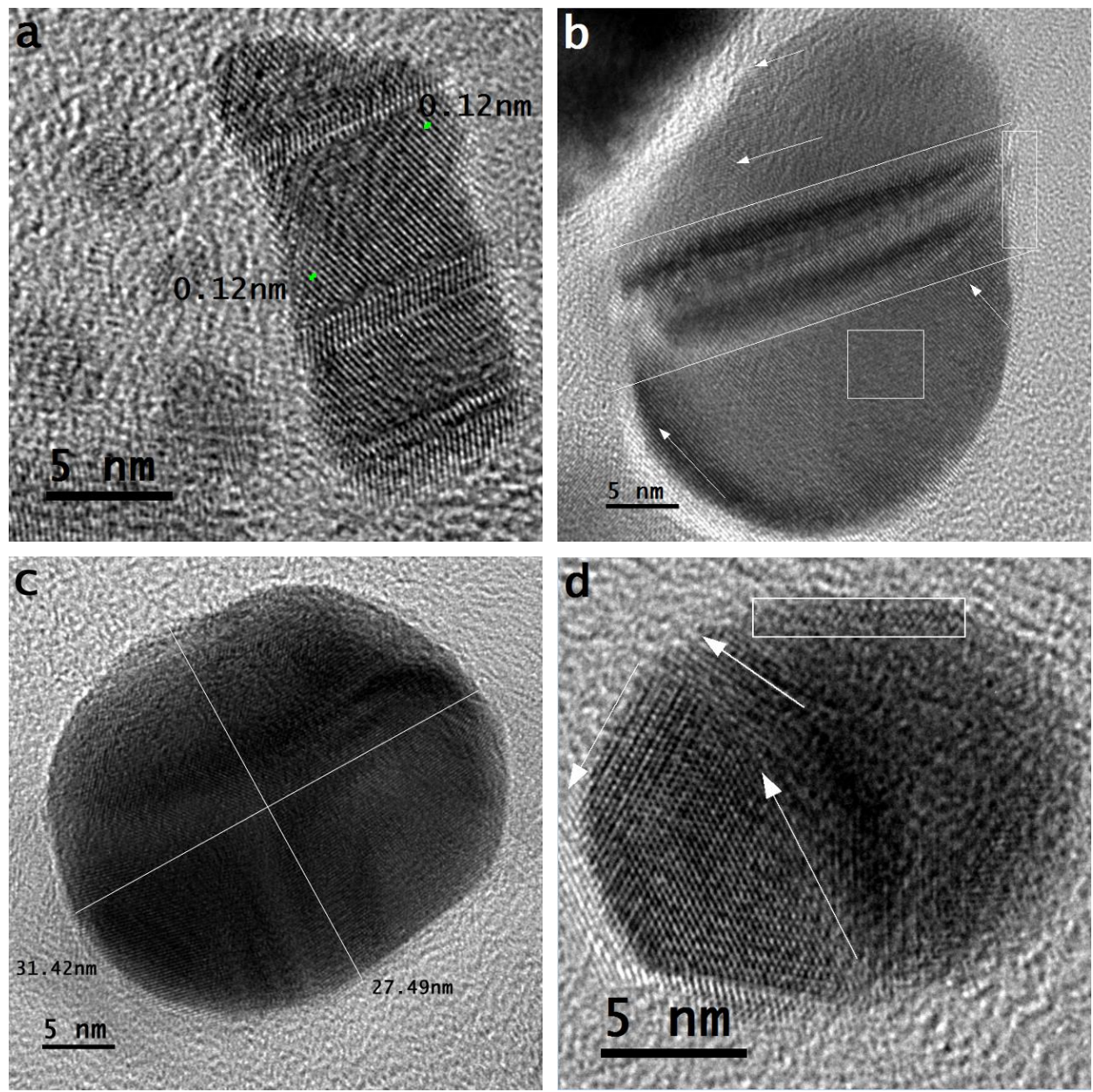

**Figure 2:** (a-d) High-resolution transmission microscope images show morphology-structure of silver tiny particles; precursor concentration: 0.60 mM, process duration: 40 minutes and argon gas flow rate: 200 sccm.

Morphology of particles resulted on processing of binary composition is shown in Figure S4 (a-d) where different features are observed as compared to silver particles. The distribution trend in particles of binary composition is also different from those of silver particles. Various bright field transmission microscope images of binary composition particles show morphology more like tetrapod shapes and individual size of particle is smaller than average size of silver particles. In Figure 3 (a), the average size of binary composition tiny particles is between 2 to 3 nm. High-resolution transmission microscope image taken from the encircled region in Figure S4 (d) is shown in Figure 3 (b) showing developing particles of binary composition through packing of tiny particles under different forces; region encircled by large circle shows the formation of smooth elements, region encircled by smaller circle shows deformation of atoms, region covered under large rectangle shows faulty structure of different stresses, region covered under smaller rectangle shows distorted structure of smooth elements, region covered under small square shows partially elongated atoms and region covered under large square shows deformation of atoms under exertion of mixed behavior force.

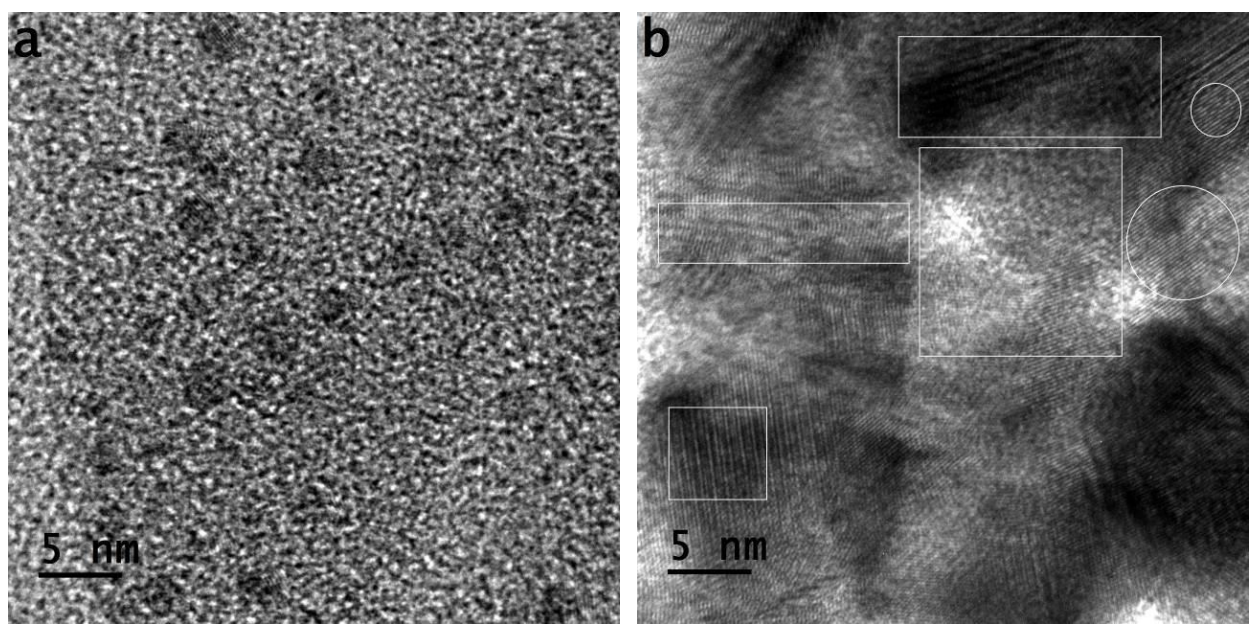

**Figure 3:** (a) High-resolution transmission microscope image of binary composition tiny particles and (b) high-resolution transmission microscope image of encircled region taken from Figure S4 (d) highlights different structural aspect at atomic level to evolve tiny particles; precursor concentration: 0.30 mM ($HAuCl_4$:$AgNO_3$ = 75%: 25 %), process duration: 20 minutes and argon gas flow rate: 100 sccm.

Figure 4 (a) shows distorted shaped gold particle. A triangular-shaped gold nanoparticle is shown in Figure 4 (b) where lengths of sides are nearly equal. Magnified high-resolution transmission microscope image was taken from the combined region of the triangular-shaped particle and distorted particle, which is shown in Figure 4 (c); in the region covered under the circle (also along the sides), deformed atoms validate their non-compact configurations where they don't have specific elongation and reveal swelling. A similar sort of atomic configuration is evident in the regions pointed out by black color arrows at middle position and at right-side where the surface is also blurred. However, in the distorted particle where the surface is pointed by the left black color arrow, atoms possess somewhat their spherical shapes where they evolve three-dimensional structure of surface format which is known as hexagonal close pack (or two-dimensional) structure. White color arrows in distorted particle (in Figure 4c) highlights several aspects of materials science; $1^{st}$ left arrow indicates more elongation of atoms, $2^{nd}$ left arrow shows stacking faults while middle arrow shows stresses, $1^{st}$ right arrow shows elongation of atoms in partially disordered structure whereas $2^{nd}$ right arrow shows deformation of atoms. In distorted particle, area covered under rectangular box shows different behaviors of deformed atoms along with packing of tiny particles. In triangular-shaped particle shown in Figure 4 (c), three important regions are labelled; atoms elongated from their centers equally under exerting force along their opposite poles because those atoms of one-dimensional arrays dealt required certain transition of their electrons (1), a large region where atoms don't elongate and retain three-dimensional structure as their electrons don't entertain the certain transition required for exertion of force (2). However, in Figure 4 (c) region (3) reveals partially deformed and partially elongated structure of atoms, which is related to blurred structure.

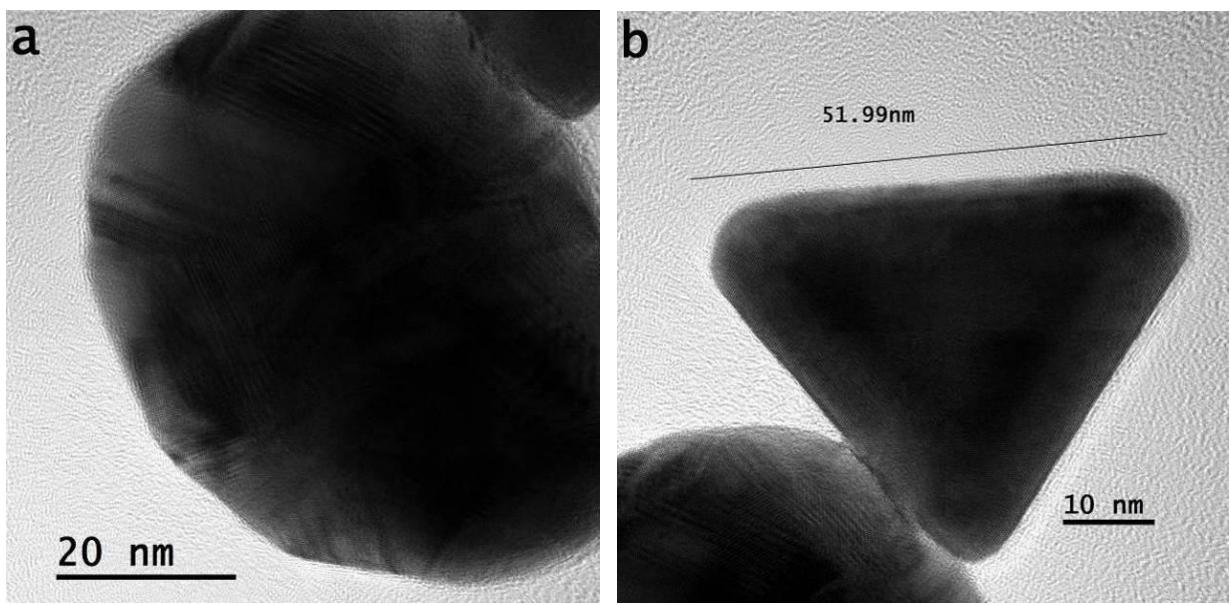

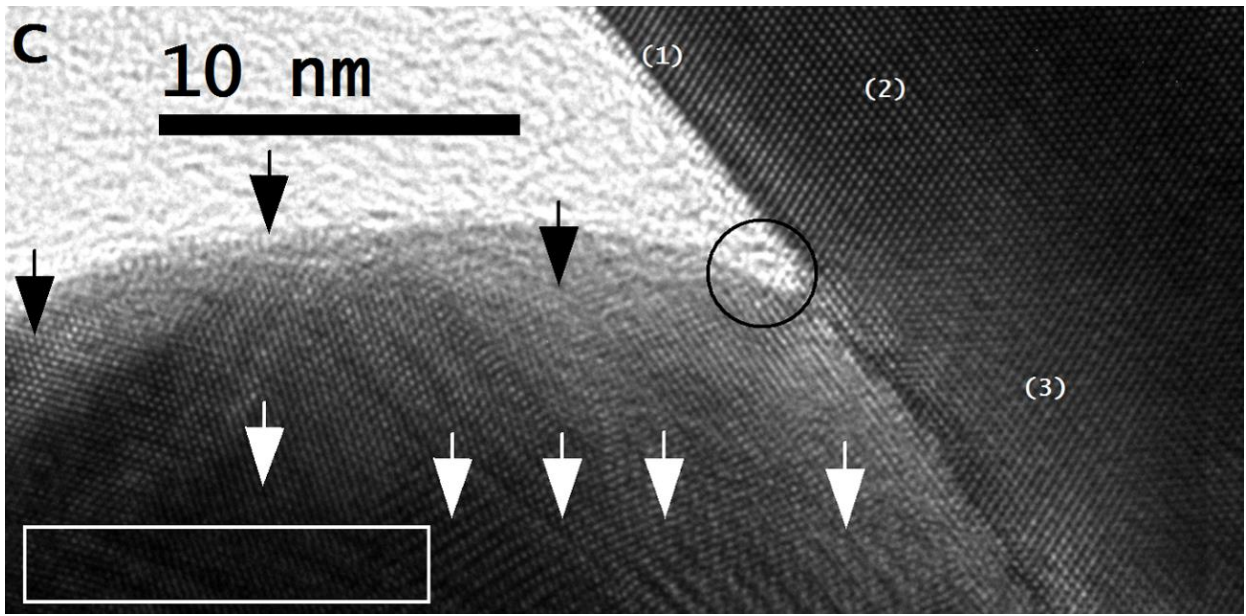

**Figure 4:** (a) Bright field transmission microscope image of gold distorted particle (b) Bright field transmission microscope image of gold geometric anisotropic shaped particle and (c) high-resolution transmission microscope image of distorted particle and geometric anisotropic shaped particle highlighting morphology-structure; precursor concentration: 0.20 mM, process duration: 5 minutes and argon gas flow rate: 100 sccm.

A high aspect ratio bar-shaped gold particle is shown in Figure 5 (a) and magnified high-resolution transmission microscope images taken from the marked regions '1' and '2' are shown on right-side where triangular-shaped tiny particles developed in structure of smooth elements resulted into packing under uniform immersing force to develop bar-shaped particle; two different regions of particle which cropped from different locations of bar-shaped particle show different orientation of packing of tiny particles as the developed structure of smooth elements show different orientation of packing. The inter-spacing distance of structure of smooth elements is approximately equal to width of a structure of smooth element (~0.12 nm) and move parallel to each other. High aspect ratio triangular-shaped gold particle is shown in Figure 5 (b) where the shape was developed in precision of an atom. The length of each side is ~655.50 nm; the shape is very thin indicating the packing of only few mono layers. In the SAPR pattern (right-side), intensity spots exhibit exact structural information of the shape where distance between any two nearest dots of intensity is ~0.24 nm and distribution of reflected photons spotted intensity spots uniformly in their printed pattern.

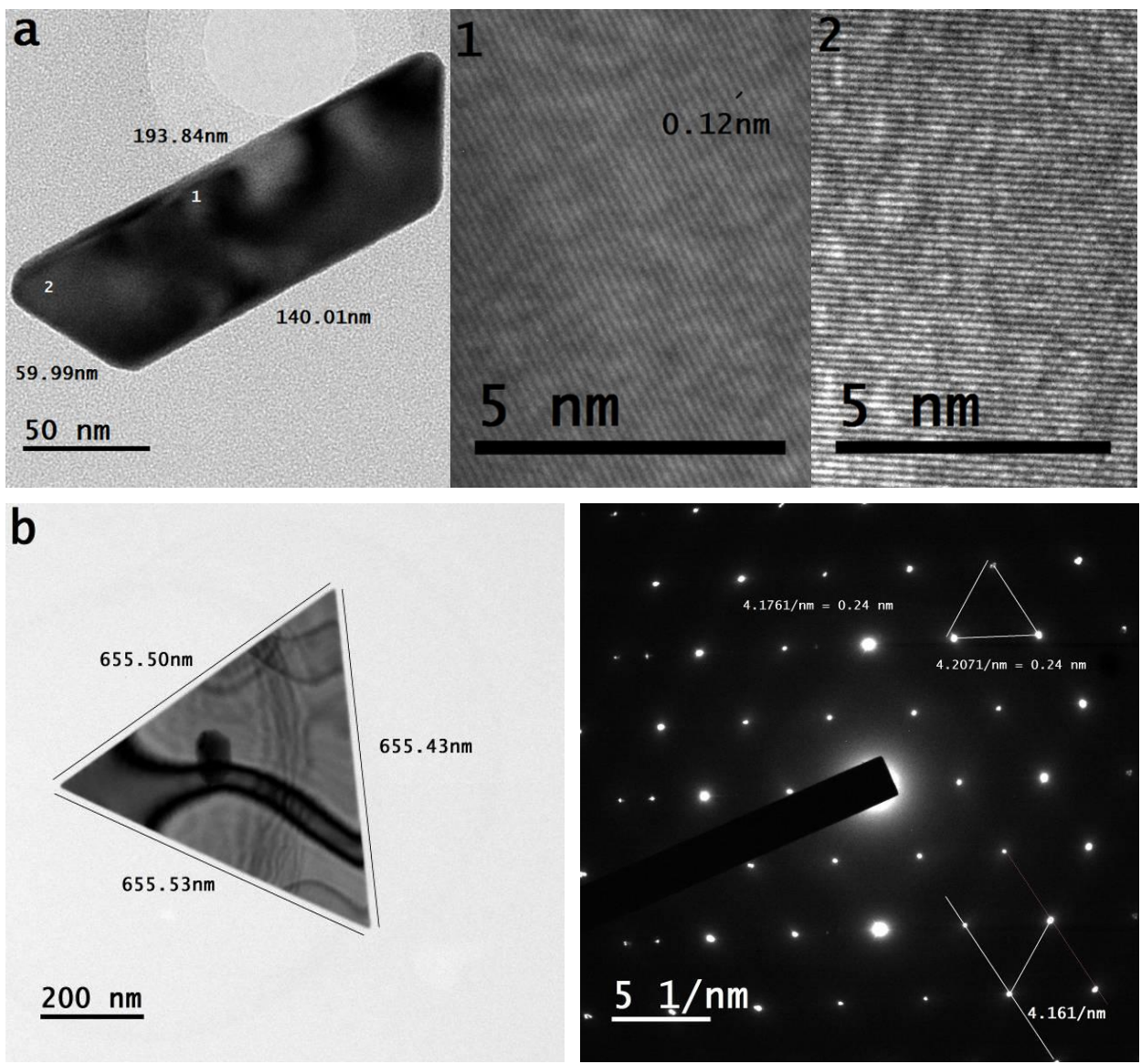

**Figure 5:** (a) Bright field transmission microscope image of bar-shaped particle (left-side) and high-magnification views of transmission microscope images taken from the marked regions '1' and '2' (right-side) and (b) bright field transmission microscope image of triangular-shaped particle (at left-side) and SAPR pattern (at right-side); precursor concentration: 0.60 mM, process duration: 5 minutes and argon gas flow rate: 100 sccm.

Formation of different tiny shaped particles has been discussed elsewhere [36]. A detailed study on the development mechanism of tiny particles shape-like equilateral triangle is given elsewhere [37-39] and similar morphology-structure of particles is shown in Figures S5 and S6 but processed under slightly varied parameters. Distorted particles of gold were also developed [Figures S5 (e-h) and Figure S6 (h)] and their detail information is given elsewhere [37-39]. Both, energy and force contribute in developing different kinds of morphology-structure tiny particles, nanoparticles and particles [39]. A detail on the formation of triangular-shaped tiny particle along with elongation of atoms is given elsewhere [40]. Gold particles of unprecedented shapes developed under tailored process parameters as discussed elsewhere [41]. Different binding mechanisms in evolving structures of solid atoms having different-dimension and different-format have been discussed elsewhere [42] where conservation forces involved to transform heat energy into binding energy under confined inter-state electron-dynamics. Atoms of electronic transition don't ionize, they either elongate or deform whereas atoms of inert gas split under the application of photonic current [43]. The phenomena of heat and photon energy have been discussed elsewhere [44] where matter at atomic scale is remained an intermediate component.

Enhanced field emission behavior of tiny grains carbon film is related to those tiny grains evolved in graphite structure and converted structure to structure of smooth elements as discussed elsewhere [45]. Under different conditions of the process, a switching morphology-structure of tiny grains, grains and crystallites in carbon film is discussed elsewhere [46]. Atomic structure of different states carbon atoms along with binding mechanism in identical state carbon atoms is discussed elsewhere [47] and content-specific growth deal certain characteristic energy when depositing films synthesized under varying chamber pressure [48]. Hard coating developed under the opposite force-energy behaviors of involved atoms while locating ground points between the mid-points of original ground points is discussed elsewhere [49]. The origins of atoms to be in gas and to be in solid state under the exceeded levitation behavior and gravitation behavior, respectively, have been discussed elsewhere [50].

In pulse-based electronphoton-solution interface process, tiny particles developed under the processing of three different solutions of precursors at slightly varied input power as given in the experimental details. However, light glow is sustained by default under each given value of pulse input power. The small difference in the input power should not turn into such a large difference in the resulted features of nanoparticles and particles, which indicate some other factors, namely, role of physical nature of metallic elements along with their affinity to precursor while processing, which commenced different dynamics of the atoms and registered in the form of different shapes of tiny particle under the same ratio of pulse OFF to ON time and so in the case of large-sized particles. As gold, silver and their binary composition hide different nature, their atoms don't detach from related precursors under the same amount of energy. Gold atoms dissociated at much faster rate under the given input power and uplifted to solution surface at uniform rate resulting into formation of tiny particles shape like equilateral triangle under same set ratio of pulse OFF to ON time as for the case silver precursor and binary composition precursor. Distribution of silver atoms at air-solution interface is appeared to be non-uniform where placed nano shape energy doesn't convert atoms of monolayer into tiny particles of triangular shape. Thus, such tiny particles don't pack under uniform forces resulting into develop the distorted particles. In binary composition, mechanisms of formation of tiny particles are even more complex and set parameters of the process don't favour the localized parameters to amalgamate atoms of binary composition into compact monolayer assembly. Under the set pulse ON/OFF time processing of silver solution and binary composition solution due to non-uniform rate of uplifting of atoms, they don't amalgamate into tiny particles having equilateral triangle shape. Again, atoms of silver and gold elements have different nature of their structure and processing their binary composition don't trigger uniform rate of amalgamation at air-solution interface that is required to design the equilateral triangular-shaped tiny particles when packets of nano shape energy were placed over the solution surface as discussed elsewhere [39]. These precursors involve different level of energy in their manufacturing and have a relation to force surrounding them in a different manner, thus, introduces a new parameter of study where nature of the precursor contributes to direct tiny particles of certain shape following by their large-sized particles. The nature of silver atoms is different as compared to nature of gold atoms. Again, when silver and gold precursors are processed under the same process parameters, they deal different behavior of processing their colloids as they require different conditions of energy and force to synergize the process in targeting tiny shaped particles. This is because of the reason that sorption activity of gold is greater than silver [51].

The uplift of silver atoms and binary composition atoms to solution surface require different scale tempo as compared to gold atoms or others depending on the nature of their atomic structure; electron (s) of the outer most ring of atoms play key role in determining the dissociation rate of metallic atoms from their precursor along with uplift to solution surface. There is a need to quantify both dissociation rate and uplifting rate of metallic atoms in all suitable materials while processing their colloids under the application of any approach. In atoms of certain elements, their behavior brings serious consequences in the nanomedicine use if the formation of their tiny particles is not properly addressed [52]. As gold precursor under suitable process conditions is enabling the monolayer assembly at solution surface where under the series of steps, it results into develop large size geometric anisotropic shapes of gold particles, which is not the case when silver solution and binary composition solution (silver and gold precursors) are processed at almost the same conditions of their process. Prescribed hcp structure is not related to two-dimensional structure but three-dimensional structure of surface format. When in the neutral state of atoms, the structure is maintained as it is, without dealing the surface format force influencing through opposite pole at centre of each atom. When atoms are in transition state, they elongate at centre to both sides under surface format force resulting into convert their one-dimensional array into structure of smooth element having nearly identical width and thickness as discussed elsewhere [40]. It may also be the case when atoms of tiny particles remain hidden from unavoidable surface format force through the brackets of companion ones where they don't permit to elongate atoms under the influence of surface format force. Nevertheless, it is convenient and worth-standing to deduce at first hand that silver atoms maintain originally different nature, then, different dissociation rate and uplift rate to solution surface along with point of transition state as compared to gold atoms. Surface format force is related to east west poles or west east poles, upper east lower west poles or lower east upper west poles and rear north right pole rear south left pole or rear north left pole rear south right pole as discussed elsewhere [42].

At pulse-based electronphoton-solution interface, impinging electron streams may deform or elongate underlying elongated atoms of tiny particles depending on the mode of impingement. Dissociated need-in atoms along with their amalgamation deal the same sorts of behavior under the process of synergy as well because the same setup and process conditions were chosen in the processing of all three different precursors.

**Conclusions:**

Under nearly identical conditions of the process while employing pulse-based electronphoton-solution interface process, the solutions of silver, binary composition (of gold and silver) and gold precursors process. Many tiny-sized particles developed have no specific geometry in the case of $AgNO_3$ precursor and $HAuCl_4.3H_2O+AgNO_3$ precursor, whereas, many tiny-sized particles developed in a geometric shape in the case of $HAuCl_4.3H_2O$ precursor. The formation process of tiny-sized particles along with developing large-sized particles in three different precursors under the same setup and process conditions result into quite a large different morphology-structure of nanoparticles and particles. Different dissociation rates of different nature atoms are because of their different force-energy behaviors and so in their precursors resulting into operate differently depsite of the fact that same process conditions opted to tap certain geometric tiny-sized particles and large-sized particles. Dissociation of silver atoms from $AgNO_3$ is not in the order of developing triangular-shaped tiny particles while the dissociation of gold atoms from the $HAuCl_4$ is in highly order not only developing triangular-shaped tiny particles but also their large-sized particles. This attribute results into develop distorted particles of silver and its binary composition, whereas, geometric aniostropic shaped particles while processing gold precursor standalone. In tiny-sized particle of no specific shape, atoms deformed instead of elongating due to mixed behavior of force, hence, in the case of their distorted particles. In the case where gold atoms retain shape, they don't deal force to elongate. Study physically shows several morphological-structural aspects of metallic colloids. Overall, at same process parameters and setup, the nature of the precursor allocates role to metallic atoms in relation to process's conditions to develop structure of tiny particles and large-sized particles of certain scope and application.

**Acknowledgements**

Mubarak Ali thanks National Science Council (now MOST) Taiwan (R.O.C.) for awarding postdoctorship: NSC-102-2811-M-032-008 (August 2013- July 2014). Authors wish to thank Mr. Chien-Jui Yeh and Dr. Kamatchi Jothiramalingam Sankaran, National

Tsing Hua University, Taiwan (R.O.C.) for their support in transmission microscopy operation. Mubarak Ali greatly appreciates useful suggestions of Dr. M. Ashraf Atta.

**Authors' biography:**

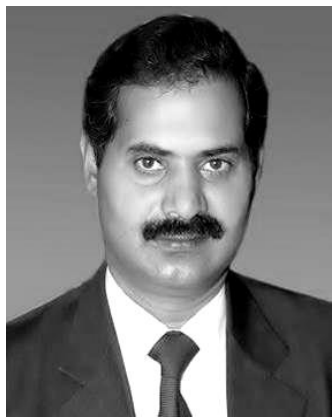

Mubarak Ali graduated from University of the Punjab with B.Sc. (Phys& Maths) in 1996 and M.Sc. Materials Science with distinction at Bahauddin Zakariya University, Multan, Pakistan (1998); thesis work completed at Quaid-i-Azam University Islamabad. He gained Ph.D. in Mechanical Engineering from Universiti Teknologi Malaysia under the award of Malaysian Technical Cooperation Programme (MTCP;2004-07) and postdoc in advanced surface technologies at Istanbul Technical University under the foreign fellowship of The Scientific and Technological Research Council of Turkey (TÜBİTAK; 2010). He completed another postdoc in the field of nanotechnology at Tamkang University Taipei (2013-2014) sponsored by National Science Council now M/o Science and Technology, Taiwan (R.O.C.). Presently, he is working as Assistant Professor on tenure track at COMSATS Institute of Information Technology, Islamabad campus, Pakistan (since May 2008) and prior to that worked as assistant director/deputy director at M/o Science & Technology (Pakistan Council of Renewable Energy Technologies, Islamabad; 2000-2008). He was invited by Institute for Materials Research (IMR), Tohoku University, Japan to deliver scientific talk on growth of synthetic diamond without seeding treatment and synthesis of tantalum carbide. He gave several scientific talks in various countries. His core area of research includes materials science, physics & nanotechnology. He was also offered the merit scholarship (for PhD study) by the Government of Pakistan but he couldn't avail. He is author of several articles published at various forums; https://scholar.google.com.pk/citations?hl=en&user=UYjvhDwAAAAJ

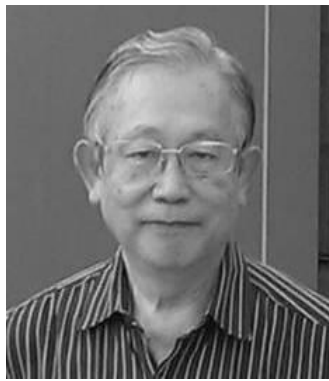

I-Nan Lin is a senior professor at Tamkang University, Taiwan. He received the Bachelor degree in physics from National Taiwan Normal University, Taiwan, M.S. from National Tsing-Hua University, Taiwan, and the Ph.D. degree in Materials Science from U. C. Berkeley in 1979, U.S.A. He worked as senior researcher in Materials Science Centre in Tsing-Hua University for several years and now is faculty in Department of Physics, Tamkang University. Professor Lin has more than 200 referred journal publications and holds top position in his university in terms of research productivity. Professor I-Nan Lin supervised several PhD and Postdoc candidates around the world. He is involved in research on the development of high conductivity diamond films and on the transmission microscopy of materials.

**Supplementary Materials:**

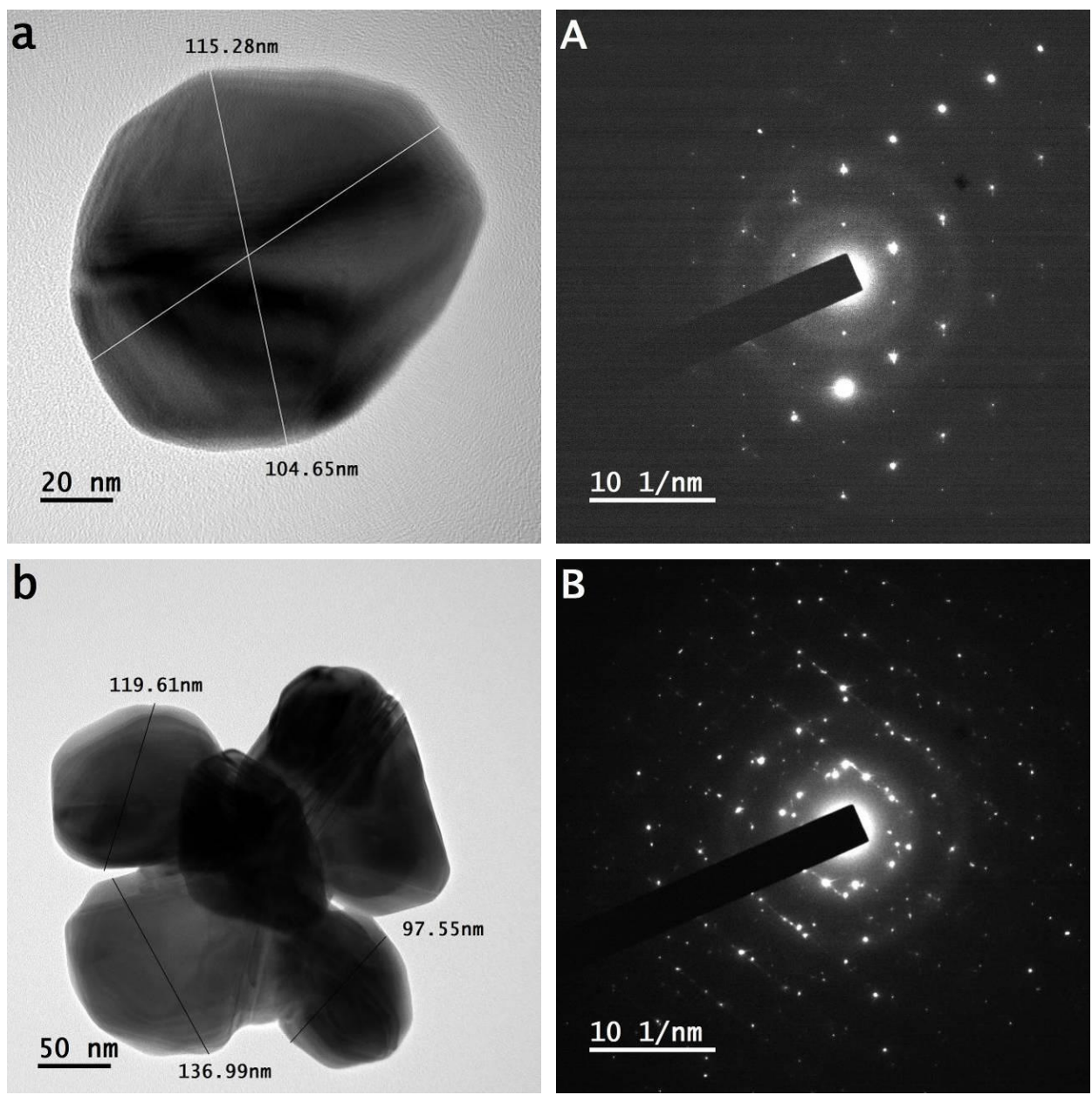

**Figure S1:** (a,b) Different bright field transmission microscope images of silver distorted particles and (A, B) SAPR patterns; precursor concentration: 0.30 mM, process duration: 20 minutes and argon gas flow rate: 100 sccm.

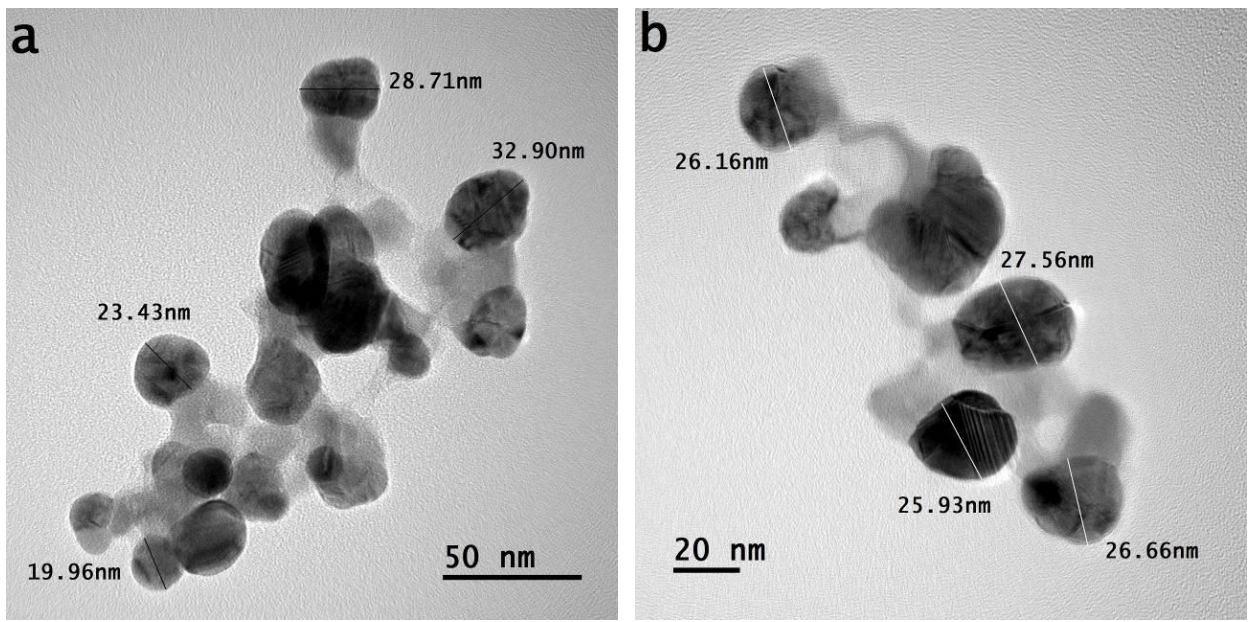

**Figure S2:** (a,b) Different bright field transmission microscope images of silver tiny particles/particles; precursor concentration: 0.30 mM, process duration: 20 minutes and argon gas flow rate: 100 sccm.

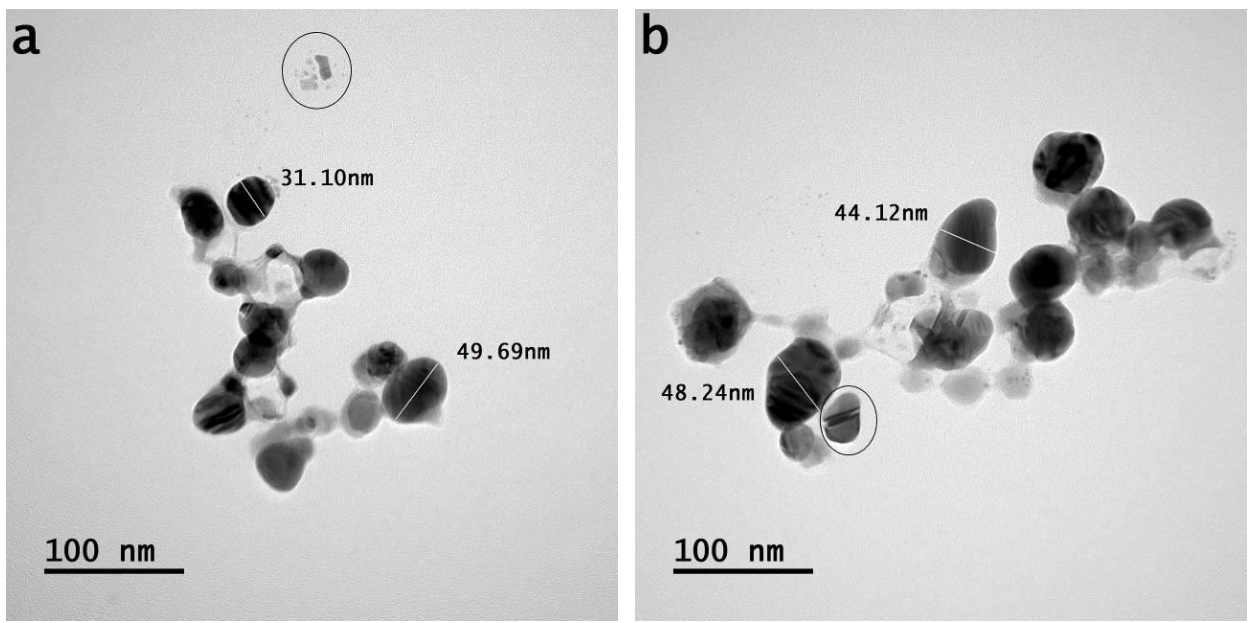

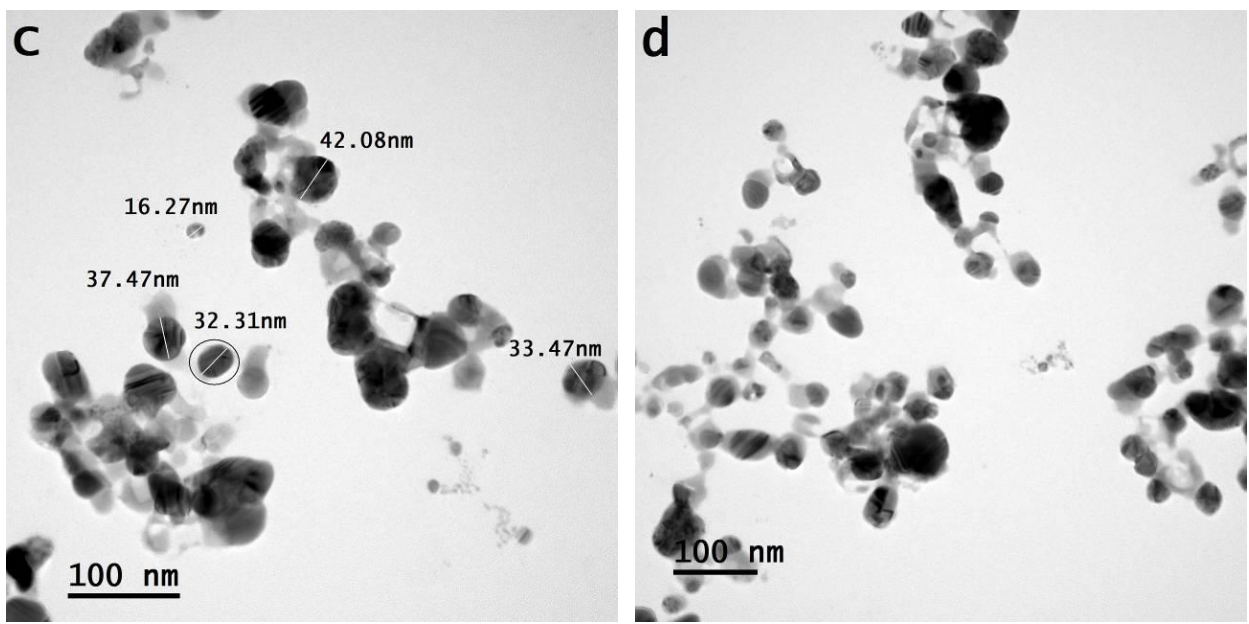

**Figure S3:** (a-d) Different bright field transmission microscope images of silver tiny particles/particles; precursor concentration: 0.60 mM, process duration: 40 minutes and argon gas flow rate: 200 sccm.

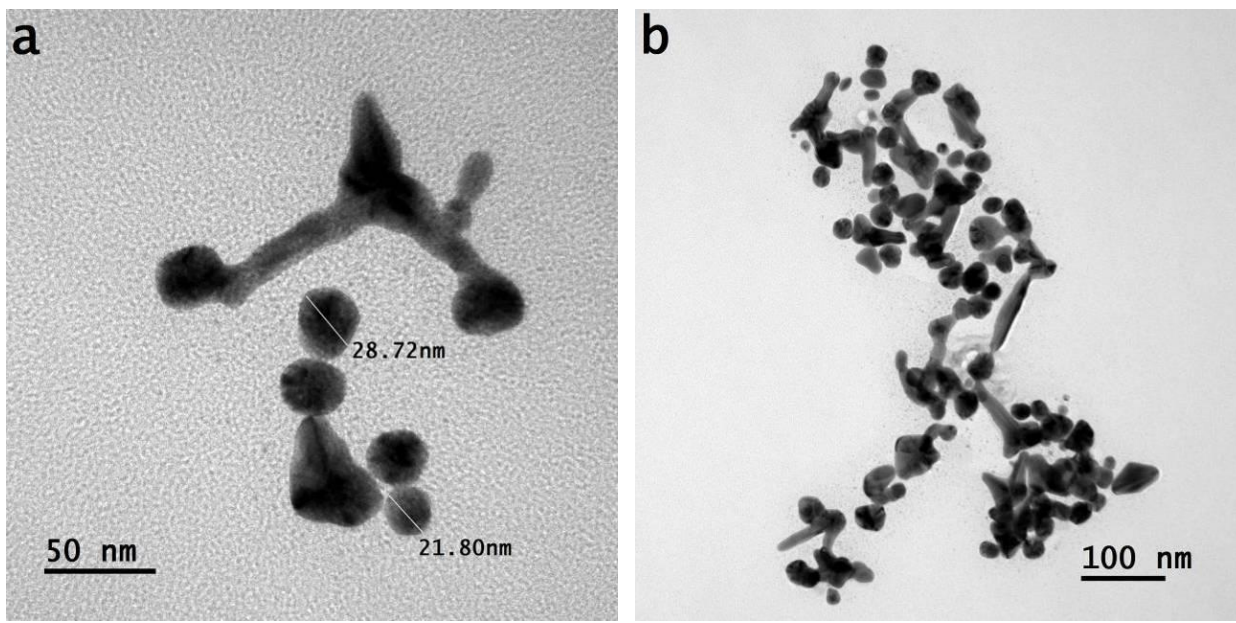

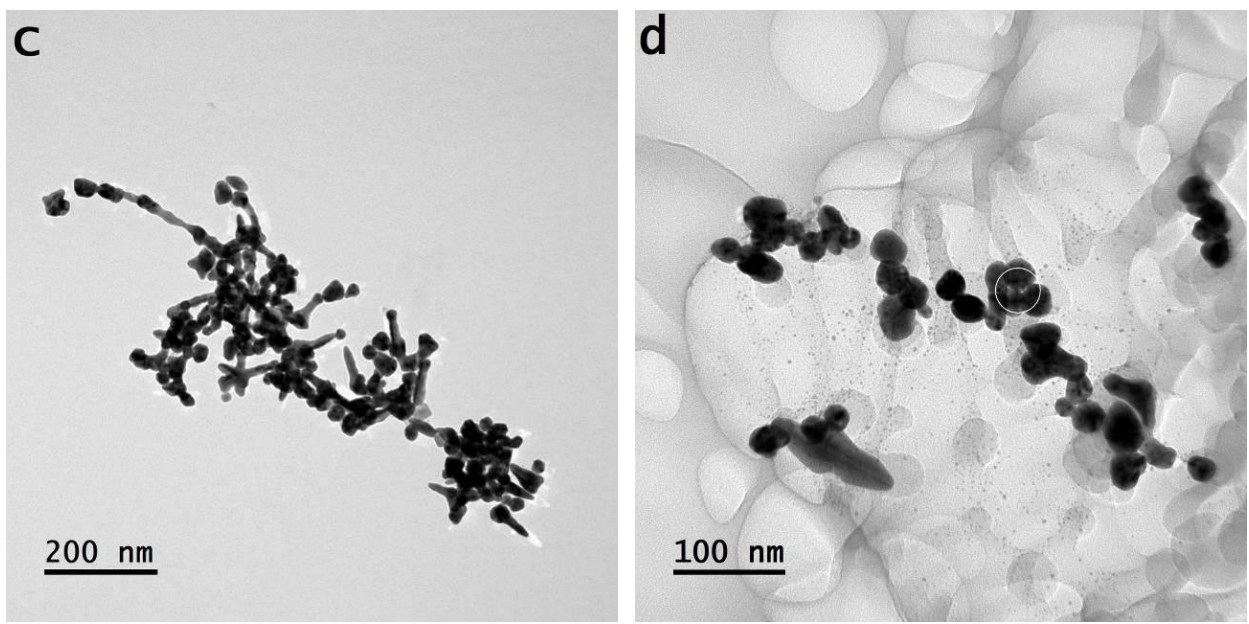

**Figure S4:** (a-d) Different bright field transmission microscope images of tiny particles/particles of binary composition; precursor concentration: 0.30 mM (Au: Ag = 75%: 25 %), process duration: 20 minutes and argon gas flow rate: 100 sccm.

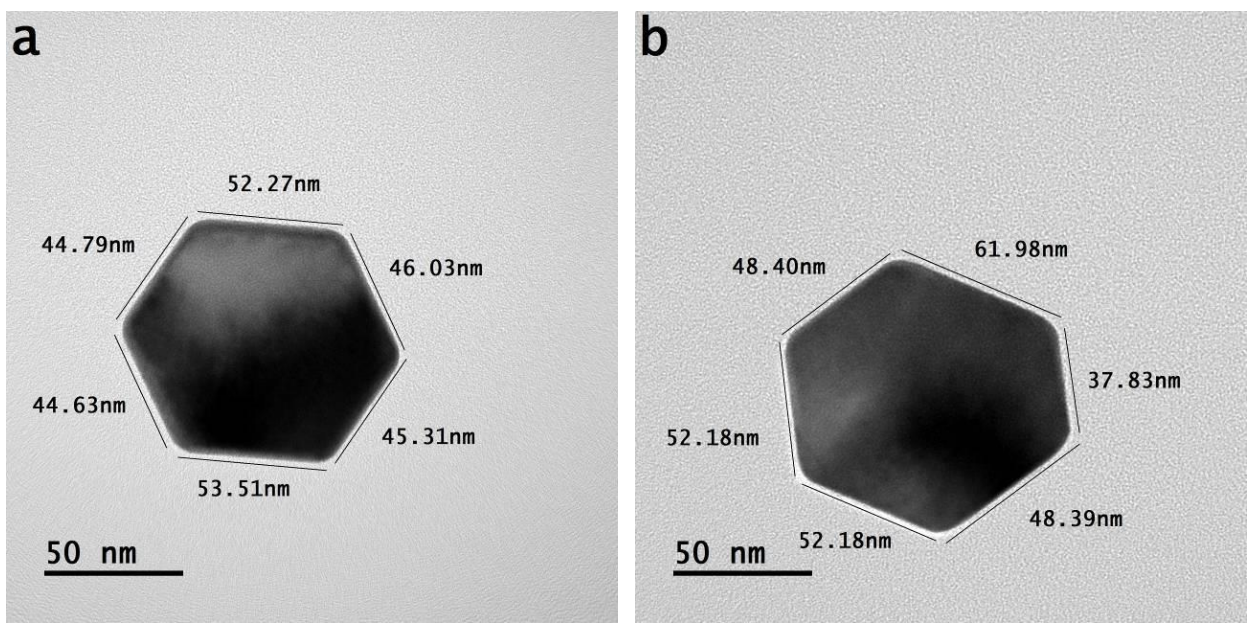

c

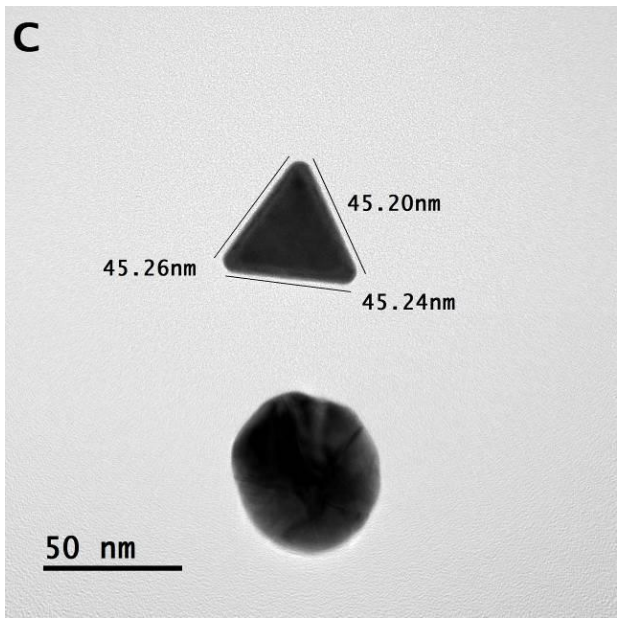

45.20nm

45.26nm

45.24nm

50 nm

d

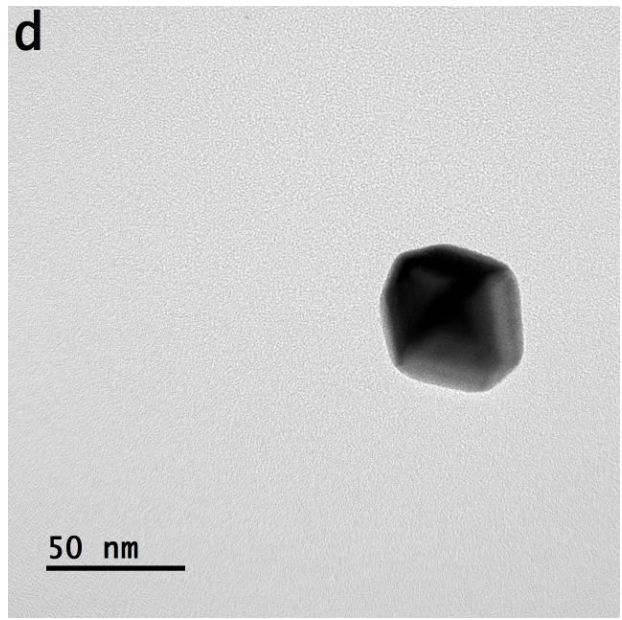

50 nm

e

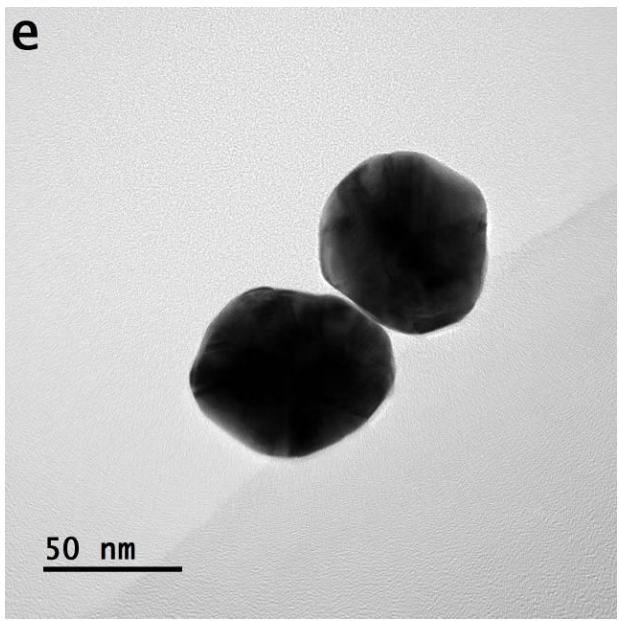

50 nm

f

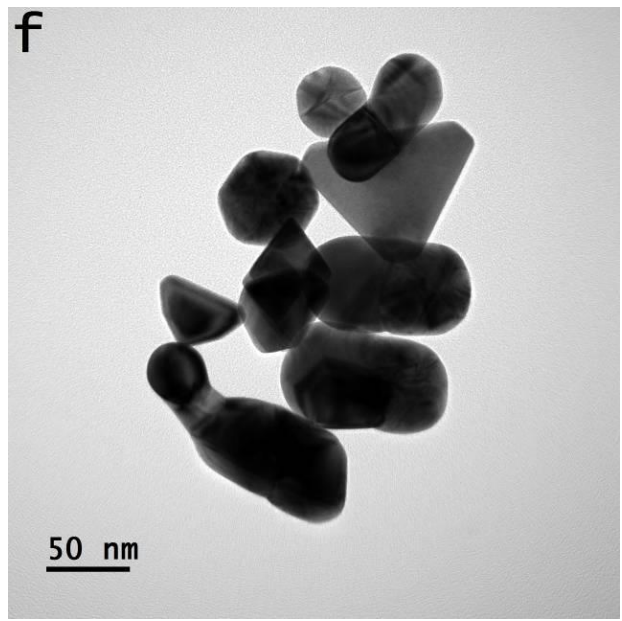

50 nm

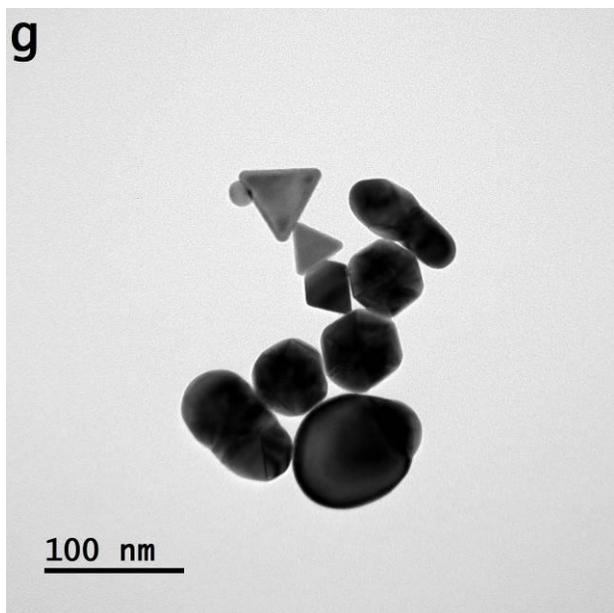

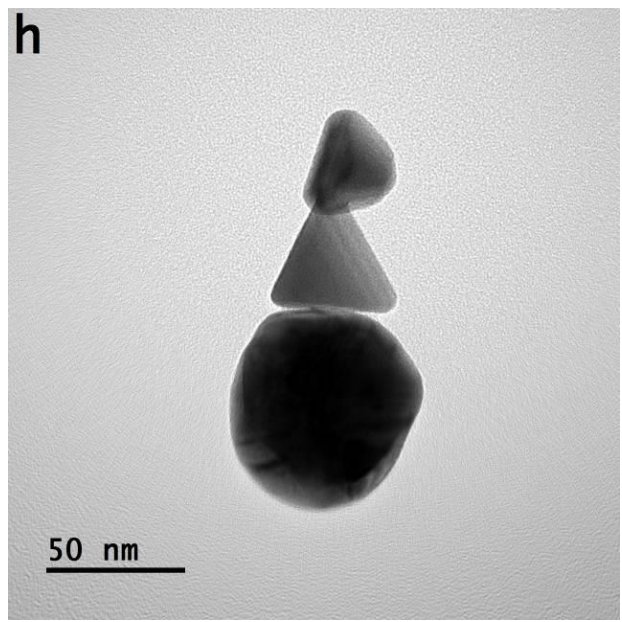

**Figure S5:** (a-h) Bright field transmission microscope images of gold geometric anisotropic shaped particles and distorted particles synthesized at precursor concentration: 0.20 mM, process duration: 5 minutes and argon gas flow rate: 100 sccm.

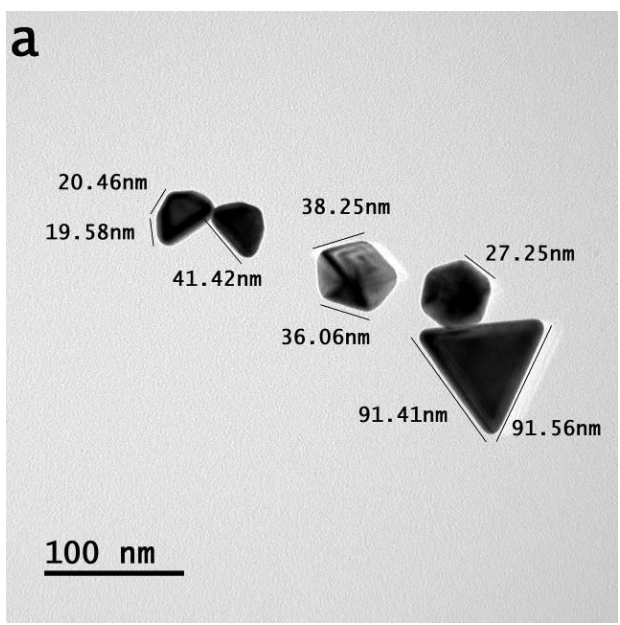

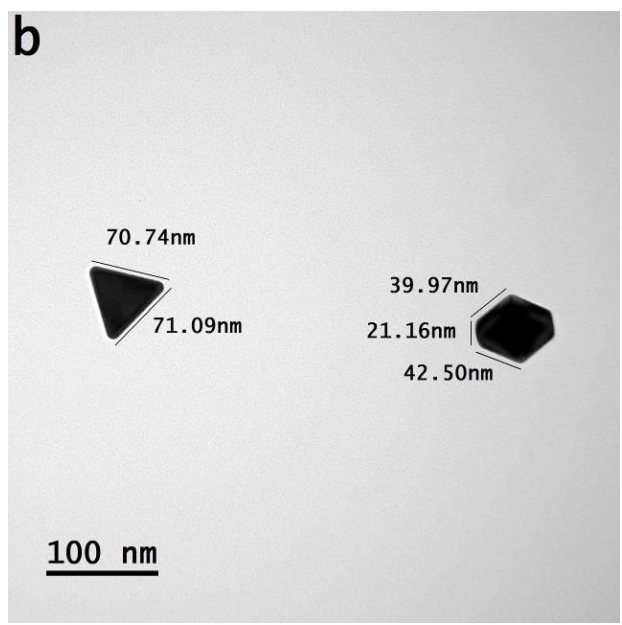

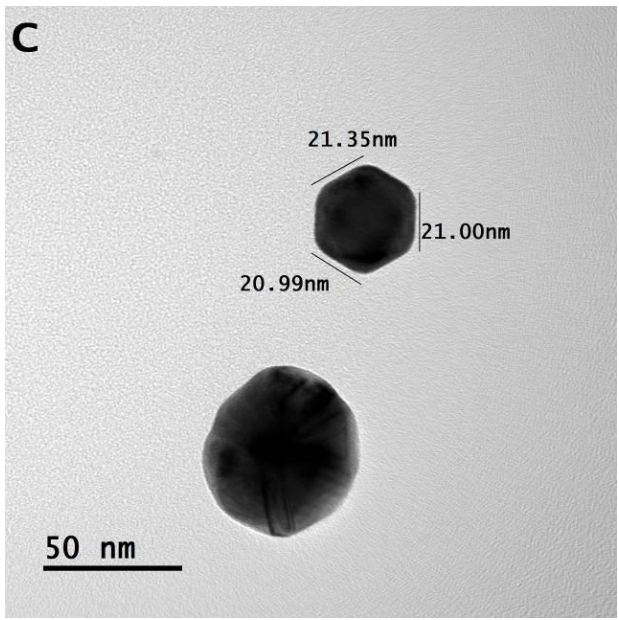

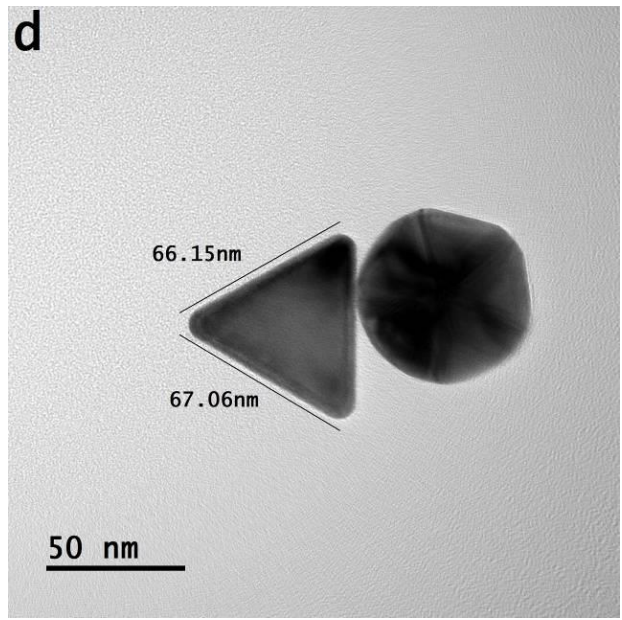

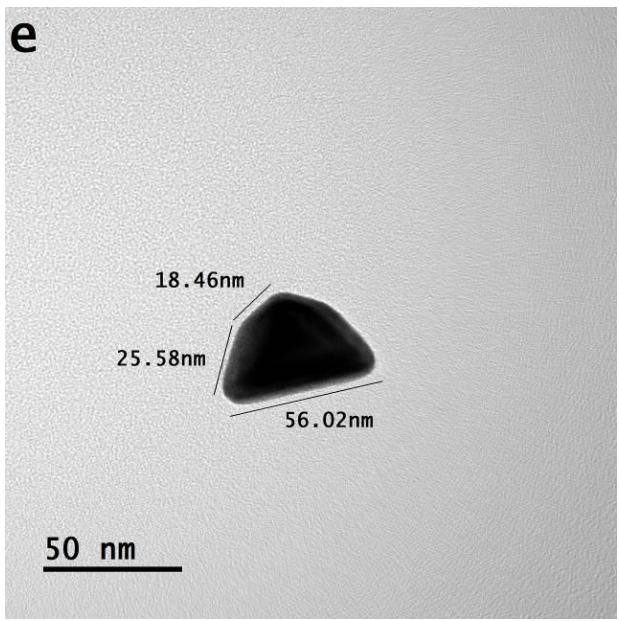

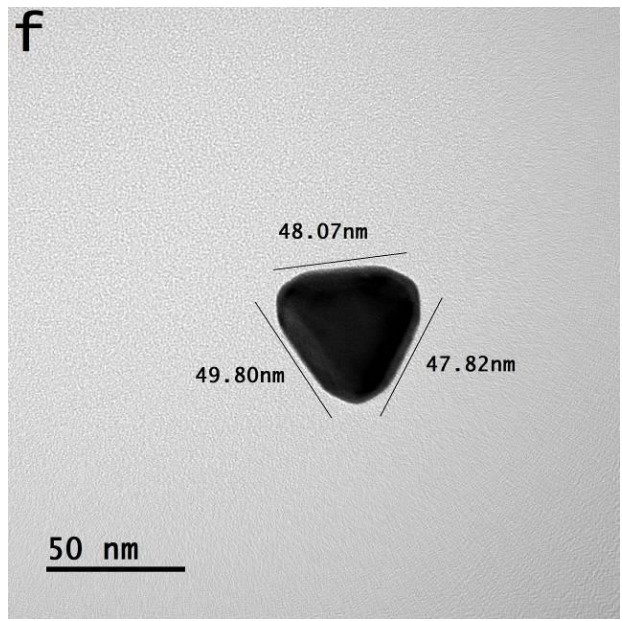

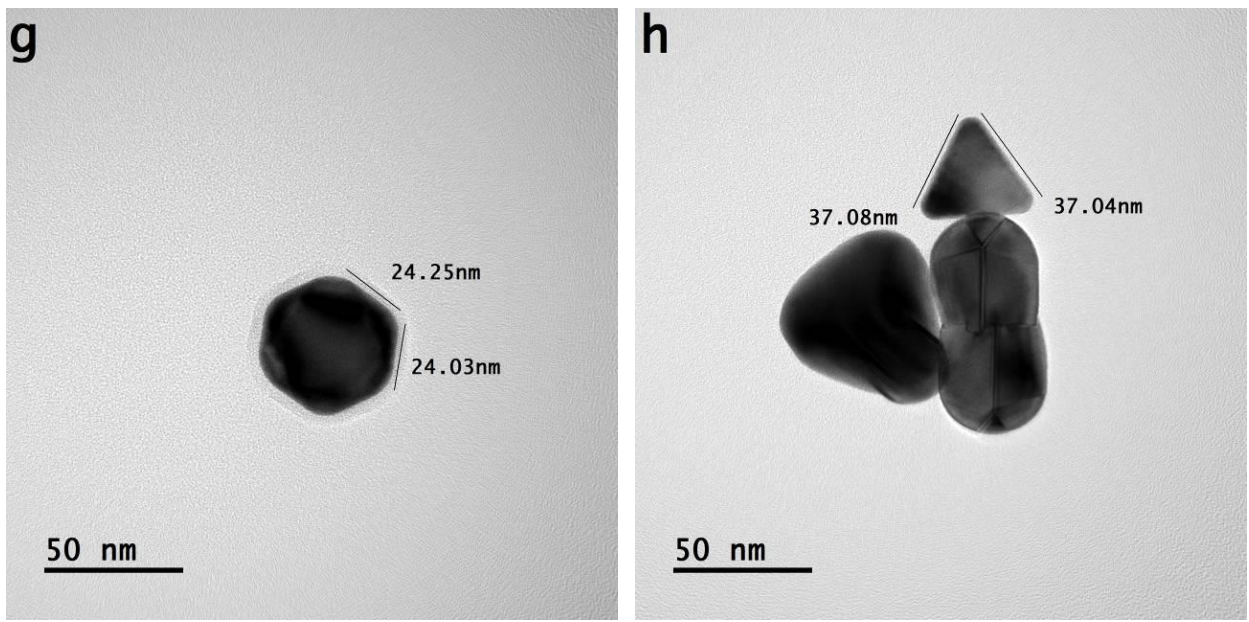

**Figure S6:** (a-h) Bright field transmission microscope images of gold geometric anisotropic shaped particles and distorted particles synthesized at precursor concentration: 0.20 mM, process duration: 10 minutes and argon gas flow rate: 100 sccm.